\documentclass[preprint,aps,amsmath,superscriptaddress,nofootinbib,tightenlines]{revtex4}
\usepackage{graphicx}
\usepackage{bm}
\usepackage{epsfig}
\usepackage{ulem}
\usepackage{color}

\def\OMIT#1{}

\newcommand{\beq}{\begin{equation}}
\newcommand{\eeq}{\end{equation}}
\newcommand{\bea}{\begin{eqnarray}}
\newcommand{\eea}{\end{eqnarray}}

\begin{document}


\title{The role of Glauber Exchange in Soft Collinear Effective Theory and 
the Balitsky-Fadin-Kuraev-Lipatov Equation}

\author{Sean Fleming\footnote{Electronic address: fleming@physics.arizona.edu}}
  \affiliation{University of Arizona, Tucson, AZ 85721, USA}
  
\date{\today\\ \vspace{1cm} }


\begin{abstract}
In soft collinear effective theory (SCET) the interaction between high energy quarks moving in opposite directions involving momentum transfer
much smaller than the center-of-mass energy is described by the Glauber interaction operator which has two-dimensional Coulomb-like behavior. Here, we determine this $n$-$\bar{n}$ collinear Glauber interaction operator and consider its renormalization properties at one loop. At this order a rapidity divergence appears which gives rise to an infrared divergent (IR) rapidity anomalous dimension commonly called the gluon Regge trajectory. We then go on to consider the forward quark scattering cross section in SCET. The emission of real soft gluons from the Glauber interaction gives rise to the Lipatov vertex. Squaring and adding the real and virtual amplitudes results in a cancelation of IR divergences, however the rapidity divergence remains. We introduce a rapidity counterterm to cancel the rapidity divergence, and derive a rapidity renormalization group equation which is the Balitsky-Fadin-Kuraev-Lipatov Equation. This connects Glauber interactions with the emergence of Regge behavior in SCET.

\end{abstract}

\maketitle

\newpage

Factorization of high-energy interactions in QCD is the systematic separation of different momentum regions into universal factors to all orders in the strong coupling constant $\alpha_{s}$. All-order proofs of factorization, which were first carried out by Collins, Soper and Sterman~\cite{Collins:1981ta,Collins:1981uk,Collins:1987pm,Collins:1989gx} rely on a set of powerful theoretical tools. Among these are: power counting, pinch analysis via the Landau equations~\cite{Landau:1959fi}, and the Coleman-Norton Theorem~\cite{Coleman:1965xm}. The Landau equations allow for the isolation of pinch singularities which, via the Coleman-Norton Theorem can be identified with long-distance (infrared) physics. Generically pinch singularities can be identified with one of three momentum regions: collinear, soft, or Glauber. In the collinear region internal propagators become collinear with external particles, and in the soft region they become soft relative to external particles. In either of these limits particles can approximately stay on their mass shell. The Glauber region, however, is special as it corresponds to off-shell modes (Glauber modes) with $k_{\perp}\gg k^{+},k^{-}$, which leads to a two-dimensional Coulomb-like interaction between and amongst collinear and soft particles~\cite{Collins:1981ta,Bodwin:1981fv}. The presence of Glauber interactions is problematic because they can destroy factorization~\cite{Bodwin:1981fv,Collins:1981tt}. Fortunately, it has been shown that for sufficiently inclusive quantities the sum over final-state cuts cancels unwanted pinches, and thereby eliminates Glauber contributions~\cite{Bodwin:1984hc,Collins:1985ue,Aybat:2008ct}. 

An alternative approach in deriving factorization is to use effective field theory (EFT). The EFT that describes the soft and collinear degrees of freedom which arise in factorization is soft collinear effective theory (SCET)~\cite{Bauer:2000ew,Bauer:2000yr,Bauer:2001ct,Bauer:2001yt}, and in Ref.~\cite{Bauer:2002nz} it was shown how the perturbative factorization theorems of QCD are reproduced in SCET. However, SCET as it was originally formulated did not include Glauber type interactions. An attempt to include Glauber interactions between collinear quarks moving in opposite directions in SCET was made Ref.~\cite{Liu:2008cc} where factorization of the Drell-Yan cross section was reconsidered. Unfortunately, this attempt did not account for the overlap between different moment regions and failed as a result . The analysis was taken up in Ref.~\cite{Bauer:2010cc} where it was concluded that ``for the exclusive Drell-Yan amplitude the correct effective theory would require Glauber modes.'' Though the authors did not consider under which circumstances the contribution from Glauber interactions cancel. In addition, a number of groups have considered the role of Glauber interactions between collinear and soft degrees of freedom in dense QCD matter~\cite{Idilbi:2008vm,DEramo:2010ak,Benzke:2012sz}. More recently, an attempt to include a Glauber interaction between two collinear particles moving in opposite directions has been presented~\cite{iainSCET10,rothsteinKITP}. 

A second, seemingly unrelated issue concerning the formulation of SCET was raised in Refs.~\cite{Donoghue:2009cq,Donoghue:2009mn}, where it was pointed out that Regge behavior appears to fall outside of the usual organizing scheme of SCET. Specifically, Regge behavior refers to the emergence of power-law behavior for scattering amplitudes. In perturbative QCD this arises out of a summation of ladder graphs which gives rise to the Balitsky-Fadin-Kuraev-Lipatov (BFKL) evolution equation~\cite{Fadin:1975cb,Balitsky:1978ic} (see Ref.~\cite{Kovchegov:2012mbw} for a very readable treatment). The solution of the leading logarithmic (LL) BFKL equation gives the total cross section for high energy scattering with just such a power law form. Clearly, as an EFT of QCD at high energy SCET needs to be able to reproduce the BFKL results. 

In this work we will show that Glauber interactions between two collinear quarks moving in opposite directions and Regge behavior are intimately connected. One cannot have one without the other. We consider the simplest possible process: two high energy quarks undergoing forward scattering. This interaction is mediated by the exchange of a gluon with momentum that has Glauber scaling (in other words a Glauber gluon): $k^{2}_{\perp}\gg k^{+}k^{-}$. We first determine the SCET operator responsible for the Glauber interaction, and then renormalize it. At one-loop a rapidity divergence appears which we treat in the formalism of Ref.~\cite{Chiu:2011qc,Chiu:2012ir}. The coefficient of the rapidity divergent term is called the gluon Regge trajectory which is infrared (IR) divergent. We then go on to consider the real emission of a soft gluon from the Glauber interaction and derive the Lipatov vertex. With these results in hand we calculate the total cross section for the forward scattering of high energy quarks. We find that at next-to-leading order in $\alpha_{s}$ this expression also has a rapidity divergence. Absorbing this rapidity divergence into a rapidity counter-term allows us to derive a rapidity RGE which is the famous BFKL equation. This then demonstrates the emergence of Regge behavior in SCET from Glauber interactions between collinear particles.

\begin{figure}
\begin{center}
\includegraphics[width=5in]{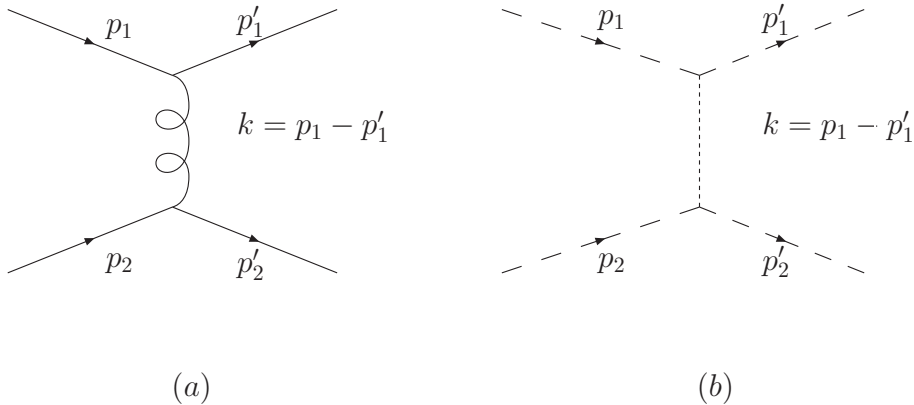}
\caption{Leading order contribution to forward quark-quark scattering at high energy: $(a)$ QCD diagram, $(b)$ SCET diagram (dashed lines indicate collinear quarks, and dotted lines Glauber gluons).}
\label{matching}
\end{center}
\end{figure}

\begin{figure}
\begin{center}
\includegraphics[width=4in]{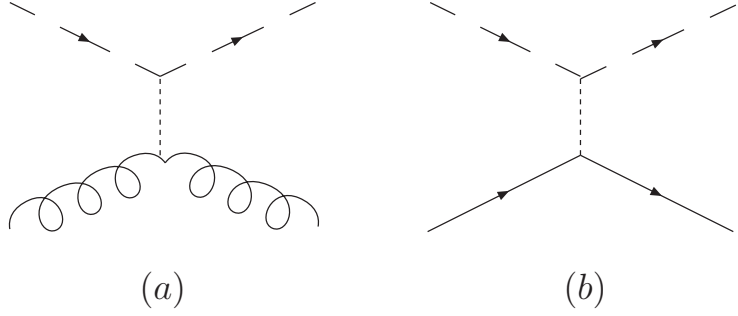}
\caption{Leading order Feynman diagrams corresponding to operators that couple collinear and soft degrees of freedom via Glauber exchange: $(a)$ collinear quark coupling to a soft gluon, $(b)$ collinear quark coupling to a soft quark (solid line). Not shown is the collinear quark coupling to a soft ghost.}
\label{csG}
\end{center}
\end{figure}

We use SCET to study the scattering of two high energy quarks moving in opposite directions $q(p_{1})+q(p_{2}) \to q(p'_{1})+q(p'_{2})$ with large invariant mass $s=(p_{1}+p_{2})^{2}$ and small momentum transfer $t = (p_{1}-p'_{1})^{2}\ll s$. We also restrict ourselves to perturbative values of $t$, where $t \gg \Lambda\sim 1\, \textrm{GeV}$. At leading order in the SCET power counting such an interaction can be described by the exchange of an off-shell gluon between the quarks, resulting in a two-dimensional Coulomb like potential in transverse momentum. To see how such an operator arises in SCET we start with QCD and match onto SCET degrees of freedom. The QCD diagram is given in Fig.~\ref{matching}$(a)$.
For the sake of matching we can take all the quarks to be massless and on-shell. In addition, the momentum $\vec{p}_{1}$ defines the $z$-axis. Then, the incoming momentum can be expressed in terms of two light-like vectors $n^{\mu}=(1,0,0,1)$ and $\bar{n}^{\mu}=(1,0,0,-1)$:
\begin{equation}
p^{\mu}_{1}= \frac{\sqrt{s}}{2} n^{\mu}\qquad p^{\mu}_{2}= \frac{\sqrt{s}}{2} \bar{n}^{\mu} \,. 
\end{equation}
The outgoing momentum can be expressed in a Sudakov decomposed form as well:
\begin{eqnarray}
&& p'^{\mu}_{1}= \frac{1}{2}(\sqrt{s}-\bar{n}\cdot k)\, n^{\mu}- \frac{1}{2}  n\cdot k \, \bar{n}^{\mu}-k^{\mu}_{\perp} \\
&& p'^{\mu}_{2}=\frac{1}{2}\bar{n}\cdot k \, n^{\mu} + \frac{1}{2}(\sqrt{s}+n\cdot k)\,  \bar{n}^{\mu}+k^{\mu}_{\perp} \,. \nonumber
\end{eqnarray}
The outgoing quarks are taken to be on-shell so they must have
\begin{equation}
n\cdot k = \frac{\vec{k}^{2}_{\perp}}{\sqrt{s}-\bar{n}\cdot k} \qquad \bar{n}\cdot k=\frac{\vec{k}^{2}_{\perp}}{\sqrt{s}+n\cdot k}\,.
\end{equation}
In the forward region we have $k^{2}= n\cdot k\, \bar{n}\cdot k+\vec{k}^{2}_{\perp}=t$  so that $k^{\mu}_{\perp}\sim \sqrt{t}$ and the above equation  implies $n\cdot k \sim  \bar{n}\cdot k \sim t/\sqrt{s} \ll k^{\mu}_{\perp}$. In this region the out-going momenta reduce to 
\begin{equation}
 p'^{\mu}_{1}\approx \frac{\sqrt{s}}{2}\, n^{\mu}+ \frac{\vec{k}^{2}_{\perp}}{2}  \bar{n}^{\mu}-k^{\mu}_{\perp}
\qquad 
p'^{\mu}_{2}\approx +\frac{\vec{k}^{2}_{\perp}}{2} \, n^{\mu} + \frac{\sqrt{s}}{2}\,  \bar{n}^{\mu}+k^{\mu}_{\perp} \,,
\end{equation}
where $k^{2}\approx -\vec{k}^{2}_{\perp}$. We carry out the matching depicted in Fig.~\ref{matching} by expanding the QCD amplitude in the forward region 
\begin{equation}
\label{treematching}
{\cal A}_{QCD} = -\frac{g^{2}}{\vec{k}^{2}_{\perp}} \bar{u}(p'_{1})T^{a}\gamma^{\mu}u(p_{1}) 
\bar{u}(p'_{2}) T^{a}\gamma_{\mu}u(p_{2})\approx
-\frac{n\cdot\bar{n} \, g^{2}}{\vec{k}^{2}_{\perp}} \bar{\xi}_{n}T^{a}\frac{\slash \!\!\! \bar{n}}{2}\xi_{n}
\bar{\xi}_{\bar{n}}  T^{a}\frac{\slash \!\!\! n}{2}\xi_{\bar{n}}
\,,
\end{equation}
where $\xi_{n}$ and $\xi_{\bar{n}}$ are the high-energy limit of the QCD spinors for quarks moving in the $n^{\mu}$ and $\bar{n}^{\mu}$ direction respectively. This amplitude is reproduced by the SCET operator first derived in Ref.~\cite{iainSCET10}
\begin{equation}
{\cal O}^{n\bar{n}}_{G}= -\frac{2 \, g^{2}}{\vec{k}^{2}_{\perp}} \bar{\xi}_{p'_{1}, n}T^{a}\frac{\slash \!\!\! \bar{n}}{2}\xi_{p_{1},n}
\bar{\xi}_{p'_{2},\bar{n}}  T^{a}\frac{\slash \!\!\! n}{2}\xi_{p_{2},\bar{n}}\,,
\end{equation}
where $\xi_{p_{1},n}$ and $\xi_{p_{2},\bar{n}}$ are SCET quark fields. This operator is not gauge invariant under separate gauge transformations in the $n$ and $\bar{n}$ sectors, but can be made so by adding the appropriate SCET collinear Wilson lines~\cite{Bauer:2001ct}
\begin{equation}
W_{n}= \sum_{\textrm{\small perms}} \textrm{exp}\bigg( -\frac{g}{\bar{n}\cdot {\cal P}} \bar{n}\cdot A_{q,n} \bigg)\qquad\textrm{and}\qquad
W_{\bar{n}}= \sum_{\textrm{\small perms}} \textrm{exp}\bigg( -\frac{g}{n\cdot {\cal P}} n\cdot A_{q,\bar{n}} \bigg)\,.
\end{equation}
In addition, soft gluons with momentum that scales as $k^{\mu}_{s}\sim \sqrt{t}$ can be radiated from the collinear quarks. While such an interaction puts the collinear quark off-shell, it is order one in the power counting and must be summed into a soft Wilson line~\cite{Bauer:2001yt}
\begin{equation}
S_{n}=  \sum_{\textrm{\small perms}} \textrm{exp} \bigg(\frac{-g}{n\cdot {\cal P}} n\cdot A_{s,q}\bigg) \qquad 
S_{\bar{n}}=  \sum_{\textrm{\small perms}} \textrm{exp} \bigg(\frac{-g}{\bar{n}\cdot {\cal P}} \bar{n}\cdot A_{s,q}\bigg)\,.
\end{equation}
Including both  collinear and soft Wilson liens we arrive at the $n$-$\bar{n}$ collinear Glauber operator
\begin{equation}
\label{ccGop}
{\cal O}^{n\bar{n}}_{G}= -8 \pi \, \alpha_{s}(\mu)\,
\bar{\xi}_{p'_{2},\bar{n}} W_{\bar{n}} Y^{\dagger}_{\bar{n}}T^{a}\frac{\slash \!\!\! n}{2}Y_{\bar{n}}W^{\dagger}_{\bar{n}}\xi_{p_{2},\bar{n}}
\frac{1}{\vec{{\cal P}}^{2}_{\perp}}
\bar{\xi}_{p'_{1}, n} W_{n}Y^{\dagger}_{n}T^{a}\frac{\slash \!\!\! \bar{n}}{2}Y_{n}W^{\dagger}_{n}\xi_{p_{1},n}\,.
\end{equation}
There are also collinear-soft Glauber operators~\cite{Idilbi:2008vm,DEramo:2010ak,Benzke:2012sz} with leading order Feynman diagrams depicted in Fig.~\ref{csG} (not shown is the coupling to a soft ghost). However, we do not need these operators here. 

\begin{figure}
\begin{center}
\includegraphics[width=4in]{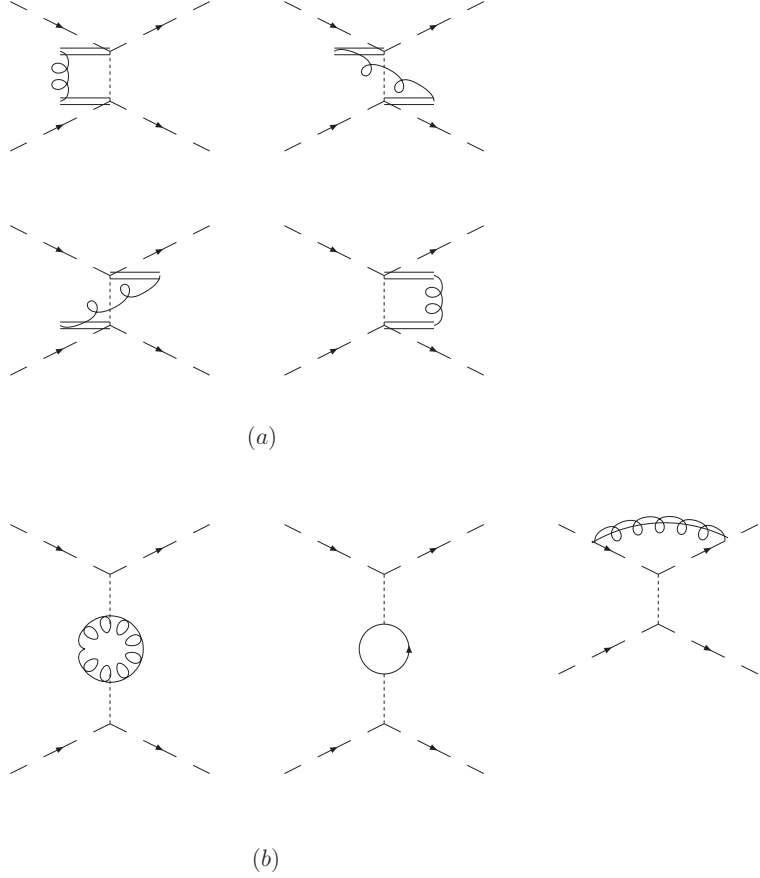}
\caption{One loop Feynman diagrams contributing to the renormalization of ${\cal O}^{n\bar{n}}_{G}$. The double line in the diagrams in $(a)$ indicate soft gluon emission from a Wilson line. These diagrams have a rapidity divergence which gives the gluon Regge trajectory. The diagrams in $(b)$ have no rapidity divergence, but have UV divergences. The first two diagrams involve soft gluons and soft quarks (the soft-ghost loop diagram is not shown), and the UV divergence in these diagrams is cancelled by a soft Lagrangian counter-term. The last diagram involves the exchange of a collinear gluon (spring with a line) and the UV divergence is cancelled by a collinear Lagrangian counter-term.}
\label{oneloop}
\end{center}
\end{figure}

Next, we renormalize the operator in Eq.~(\ref{ccGop}). The diagrams that contribute are shown in Fig.~\ref{oneloop}. The double lines in the diagrams in $(a)$ indicate that a soft gluon is emitted from one of the soft Wilson lines. These diagrams are ultraviolet (UV) finite, but contain a rapidity divergence. The diagrams in $(b)$ are UV divergent but do not have a rapidity divergence. The UV divergent terms in these diagrams are cancelled by either a soft Lagrangian counter-term [first two diagrams in $(b)$ plus a ghost-loop that is not shown], or a collinear Lagrangian counter-term [last diagram in $(b)$]. 

The physics of interest is  associated with the rapidity divergence, so we will focus on the diagrams in Fig~\ref{oneloop}$(a)$. The sum of these four diagrams gives
\begin{equation}
\label{loop}
{\cal A} = -8\pi \alpha_{s}(\mu)\, \bar{\xi}_{n}T^{a}\frac{\slash \!\!\! \bar{n}}{2}\xi_{n}
\bar{\xi}_{\bar{n}}  T^{a}\frac{\slash \!\!\! n}{2}\xi_{\bar{n}}\bigg[i N_{c} \alpha_{s}(\mu) {\cal I}(\vec{k}_{\perp}) \bigg]\,,
\end{equation}
where 
\begin{equation}
{\cal I}(\vec{k}_{\perp}) = \int \frac{d q^{-}}{q^{-}} \, \int \frac{d^{2}q_{\perp}}{(2\pi)^{2}}\frac{1}{\vec{q}^{\,\,2}_{\perp}}\frac{1}{(\vec{q}+\vec{k})^{2}_{\perp}} \,.
\end{equation}
In obtaining the expression in Eq.~(\ref{loop}) a symmetry factor of one-half needs to be included as the first two diagrams in Fig~\ref{oneloop}$(a)$ are identical to the second two diagrams.
The integral over $q^{-}$ results in a rapidity divergence, while the integral over $q_{\perp}$, which in the literature is called the gluon Regge trajectory, contains IR divergences. To evaluate this integral we will need to introduce regulators for both types of divergences. Here we will regulate the rapidity divergence using the methods developed in Ref.~\cite{Chiu:2011qc,Chiu:2012ir}, and use a gluon mass (or dimensional regularization) to regulate IR divergences. With these modifications the integral above becomes
\begin{eqnarray}
{\cal I}(\vec{k}_{\perp}) &=&\nu^{2\eta} w(\nu)^{2 } \int \frac{d^{4}q}{(2\pi)^{4}}\frac{1}{q^-}\frac{1}{q^+}\frac{(q^3)^{-2\eta}}{q^2-m^2_g}\frac{1}{(\vec{q}+\vec{k})^{2}_{\perp}+m^2_g} \\
&=& -i\frac{\nu^{2\eta}w(\nu)^{2 }}{\eta} \frac{\Gamma\big(\frac{1}{2}-\eta\big)\Gamma(1+\eta)}{(4\pi)^2\sqrt{\pi}}\frac{1}{(k^2_\perp)^{1+\eta}}\int^1_0 dx \frac{x^\eta}{[x(1-x)+m^2_g/k^2_\perp]^{1+\eta}}\nonumber\\
&\approx& \frac{-2 i}{(4 \pi)^2}\frac{w(\nu)^{2 }}{\vec{k}^2_\perp}\bigg[ \frac{1}{\eta}\ln\bigg(\frac{\vec{k}^2_\perp}{m^2_g}\bigg)+ \ln\bigg(\frac{\vec{k}^2_\perp}{4 \nu}\bigg)\ln\bigg(\frac{\vec{k}^2_\perp}{m^2_g}\bigg)-\frac{1}{4}\ln^2\bigg(\frac{\vec{k}^2_\perp}{m^2_g}\bigg)+ i \pi\ln\bigg(\frac{\vec{k}^2_\perp}{m^2_g}\bigg)\bigg] 
\nonumber \,.
\end{eqnarray}
For completeness we also give an expression for ${\cal I}(\vec{k}_{\perp})$
regulating the IR divergences with dimensional regularization:
\begin{eqnarray}
{\cal I}(\vec{k}_{\perp}) &=& -i(4 \pi \mu^{2})^{\epsilon} \frac{\nu^{2\eta}w(\nu)^{2 }}{\eta} \frac{\Gamma\big(\frac{1}{2}-\eta\big)\Gamma(1+\eta+\epsilon)}{(4\pi)^2\sqrt{\pi}}\frac{1}{(k^2_\perp)^{1+\eta+\epsilon}}\frac{\Gamma(-\epsilon)\Gamma(-\eta-\epsilon)}{\Gamma(-\eta-2 \epsilon)} \\
&\approx& \frac{-2 i}{(4 \pi)^2}\frac{w(\nu)^{2 }}{\vec{k}^2_\perp}\bigg\{  \frac{\Gamma(-\epsilon)}{\eta}\bigg(\frac{\bar{\mu}^2 e^{\gamma_{E}}}{\vec{k}^2_\perp}\bigg)^\epsilon\,\frac{\Gamma(1+\epsilon)\Gamma(1-\epsilon)}{\Gamma(1-2\epsilon)}+\frac{1}{2\epsilon^2}\nonumber \\
&& +\frac{1}{2\epsilon}\bigg[ \ln\bigg(\frac{\bar{\mu}^2}{4 \nu^2}\bigg)+\ln\bigg(\frac{\vec{k}^2_\perp}{4 \nu^2}\bigg)\bigg]
+ \frac{1}{4}\ln^2\bigg(\frac{\vec{k}^2_\perp}{\bar{\mu}^2}\bigg) - \ln\bigg(\frac{\vec{k}^2_\perp}{4\nu^2}\bigg)\ln\bigg(\frac{\vec{k}^2_\perp}{\bar{\mu}^2}\bigg)
-\frac{\pi^2}{24}\bigg\}\,,\nonumber
\end{eqnarray}
where $\bar{\mu}^2 = 4 \pi \mu^2e^{-\gamma}$.
The rapidity divergence corresponds to the term that diverges as $\eta \to 0$. This rapidity pole must be subtracted by a rapidity counter-term. However, as the rapidity divergent term contains IR divergences a sensible rapidity RGE can not be derived. This issue is fixed if we consider forward scattering and include real emission diagrams. 

\begin{figure}
\begin{center}
\includegraphics[width=3.5in]{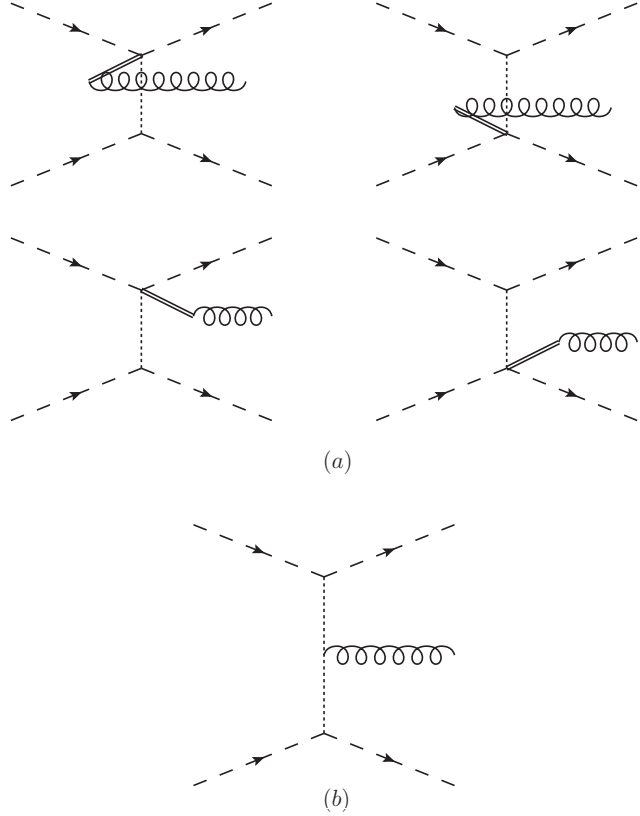}
\caption{Real emission of soft gluons from the $n$-$\bar{n}$ Glauber interaction: (a) emission from the soft Wilson lines, (b) emission from the Glauber gluon.}
\label{real}
\end{center}
\end{figure}

The emission of a real soft gluon can occur from any of the soft Wilson lines as shown in Fig.~\ref{real}(a) or from the exchanged Glauber gluon as shown in Fig.~\ref{real}(b).
The amplitude for the sum of the  four diagrams in Fig.~\ref{real}(a) is 
\begin{equation}
\sum_{i=1}^4 {\cal A}^i_\textrm{\small real} = -2 \, g^{2}\frac{1}{\vec{k}^{2}_{\perp}} \frac{1}{\vec{k}^{'2}_{\perp}} \bar{\xi}_{n}T^{a}\frac{\slash \!\!\! \bar{n}}{2}\xi_{n}
\bar{\xi}_{\bar{n}}  T^{b}\frac{\slash \!\!\! n}{2}\xi_{\bar{n}}(-i g f^{abc})\bigg(\frac{n^\alpha}{n\cdot k'} \vec{k}^2_\perp+\frac{\bar{n}^\alpha}{\bar{n}\cdot k} \vec{k}^{'2}_\perp\bigg)
\end{equation}
and the amplitude for the diagram in Fig.~\ref{real}(b) is
\begin{equation}
{\cal A}^5_\textrm{\small real}=-2 \, g^{2}\frac{1}{\vec{k}^{2}_{\perp}} \frac{1}{\vec{k}^{'2}_{\perp}} \bar{\xi}_{n}T^{a}\frac{\slash \!\!\! \bar{n}}{2}\xi_{n}
\bar{\xi}_{\bar{n}}  T^{b}\frac{\slash \!\!\! n}{2}\xi_{\bar{n}}(i g f^{abc})\bigg( k^\alpha_\perp+k^{'\alpha}_\perp -\frac{1}{2}\bar{n}^\alpha n\cdot k'
-\frac{1}{2}n^\alpha \bar{n}\cdot k\bigg)
\,,
\end{equation}
where the soft gluon momentum is $q^\mu = k^\mu-k^{'\mu}\approx \frac{1}{2}\bar{n}\cdot k n^{\mu}-\frac{1}{2}n\cdot k' \bar{n}^{\mu}+(k_{\perp}-k_{\perp}')^{\mu}$.
Adding these up we arrive at the Lipatov vertex
\begin{eqnarray}
{\cal A}_\textrm{ L}&=& -2 \, g^{2}\frac{1}{\vec{k}^{2}_{\perp}} \frac{1}{\vec{k}^{'2}_{\perp}} \bar{\xi}_{n}T^{a}\frac{\slash \!\!\! \bar{n}}{2}\xi_{n}
\bar{\xi}_{\bar{n}}  T^{b}\frac{\slash \!\!\! n}{2}\xi_{\bar{n}}\\
&& \times (i g f^{abc})\bigg( k^\alpha_\perp+k^{'\alpha}_\perp -\frac{1}{2}\bar{n}^\alpha n\cdot k'
-\frac{1}{2}n^\alpha \bar{n}\cdot k-  \frac{n^\alpha}{n\cdot k'} \vec{k}^2_\perp-\frac{\bar{n}^\alpha}{\bar{n}\cdot k} \vec{k}^{'2}_\perp\bigg)\,.\nonumber
\end{eqnarray}
This vertex is gauge invariant, as can be explicitly verified by contracting with the external gluon momentum.

Now we have all the pieces needed to calculate the quark scattering cross section in the forward region. Squaring the amplitude in Eq.~(\ref{treematching}) we obtain the tree level cross section
\begin{equation}
 \sigma^{LO}= \frac{2 \alpha^{2}_{s} C_{F}}{N_{c}} \int  \frac{d^{2}\vec{k}^{2}_{\perp}}{\vec{k}_{\perp}^{2}} \int \frac{d^{2} \vec{k}^{'2}_{\perp}}{\vec{k}_{\perp}^{'2}}\,\delta^{(2)}(\vec{k}_{\perp}-\vec{k}'_{\perp})\,.
 \end{equation}
The NLO virtual corrections give
\begin{eqnarray}
 \sigma^{NLO}_{V}&=& \frac{2 \alpha^{2}_{s} C_{F}}{N_{c}}  \int  \frac{d^{2}\vec{k}_{\perp}}{\vec{k}_{\perp}^{2}} \int \frac{d^{2} \vec{k}_{\perp}'}{\vec{k}_{\perp}^{'2}}\, \delta^{(2)}(\vec{k}_{\perp}-\vec{k}'_{\perp}) \\
 &&\qquad\times \bigg(-\frac{\alpha_{s} N_{c}}{2\pi^{2}}\bigg)\nu^{2 \eta}w(\nu)^{2}\frac{\Gamma(\eta)\Gamma\big(\frac{1}{2}-\eta\big)}{\sqrt{\pi}}\int d^{2}q_{\perp} \frac{\vec{k}_{\perp}^{2}}{\vec{q}^{\,\,2}_{\perp}}\frac{1}{[(\vec{q}_{\perp}-\vec{k}_{\perp})^{2}]^{1+\eta}} \nonumber
 \end{eqnarray}
and the NLO real corrections give
\begin{eqnarray}
 \sigma^{NLO}_{R}&=& \frac{2 \alpha^{2}_{s} C_{F}}{N_{c}}  \int  \frac{d^{2}\vec{k}_{\perp}}{\vec{k}_{\perp}^{2}} \int \frac{d^{2} \vec{k}_{\perp}'}{\vec{k}_{\perp}^{'2}}\\
&& \qquad\times \bigg(\frac{\alpha_{s} N_{c}}{\pi^{2}}\bigg)\nu^{2 \eta}w(\nu)^{2}\frac{\Gamma(\eta)\Gamma\big(\frac{1}{2}-\eta\big)}{\sqrt{\pi}}
 \int d^{2}q_{\perp}\,\frac{\delta^{(2)}(\vec{q}_{\perp}-\vec{k}'_{\perp})}{[(\vec{q}_{\perp}-\vec{k}_{\perp})^{2}]^{1+\eta}} \,.\nonumber
 \end{eqnarray}
Adding these up we arrive at an expression for the forward scattering cross section accurate to NLO
\begin{eqnarray}
\sigma&=& \frac{2 \alpha^{2}_{s} C_{F}}{N_{c}}  \int  \frac{d^{2}\vec{k}_{\perp}}{\vec{k}_{\perp}^{2}} \int \frac{d^{2} \vec{k}_{\perp}}{\vec{k}_{\perp}'^{'2}}\bigg\{ \delta^{(2)}(\vec{k}_{\perp}-\vec{k}'_{\perp})
+\bigg(\frac{\alpha_{s} N_{c}}{\pi^{2}}\bigg)\frac{\Gamma(\eta)\Gamma\big(\frac{1}{2}-\eta\big)}{\sqrt{\pi}}\nu^{2 \eta}w(\nu)^{2}
\\
&& \times\int \frac{d^{2}q_{\perp}}{[(\vec{q}_{\perp}-\vec{k}_{\perp})^{2}]^{1+\eta}}\bigg[ \delta^{(2)}(\vec{q}_{\perp}-\vec{k}'_{\perp})-\frac{\vec{k}^{2}_{\perp}}{2 \vec{q}^{\,\,2}_{\perp}}\delta^{(2)}(\vec{k}_{\perp}-\vec{k}'_{\perp})\bigg]\bigg\}\nonumber\,.
\end{eqnarray}
Expanding around $\eta = 0$ we can isolate the rapidity divergent term
\begin{eqnarray}
\label{rapdivterms}
\sigma&=& \frac{2 \alpha^{2}_{s} C_{F}}{N_{c}}  \int  \frac{d^{2}\vec{k}_{\perp}}{\vec{k}_{\perp}^{2}} \int \frac{d^{2} \vec{k}_{\perp}}{\vec{k}_{\perp}'^{'2}}\bigg\{ \delta^{(2)}(\vec{k}_{\perp}-\vec{k}'_{\perp})\\
&&+\bigg(\frac{\alpha_{s} N_{c}}{\pi^{2}}\bigg)\frac{w(\nu)^{2}}{\eta}\int \frac{d^{2}q_{\perp}}{(\vec{q}_{\perp}-\vec{k}_{\perp})^{2}}\bigg[ \delta^{(2)}(\vec{q}_{\perp}-\vec{k}'_{\perp})
-\frac{\vec{k}^{2}_{\perp}}{2 \vec{q}^{\,\,2}_{\perp}}\delta^{(2)}(\vec{k}_{\perp}-\vec{k}'_{\perp})\bigg]+\dots\bigg\}\,,\nonumber
\end{eqnarray}
where the dots represent NLO terms that are finite in the $\eta \to 0$ limit.
In order to renormalize the rapidity divergence we identify the two-dimension Dirac delta function in transverse-momentum space as the leading order vacuum matrix element of an (unknown) operator, $O_{G}^{\textrm{\tiny soft}}$, involving soft fields. Let $G(\vec{k}_{\perp}-\vec{k}'_{\perp})\equiv \langle O_{G,\textrm{\tiny soft}} \rangle$, then
\begin{eqnarray}
G(\vec{k}_{\perp}-\vec{k}'_{\perp},\nu) &=& \int d^{2}\ell_{\perp} {\cal Z}^{-1}(\vec{k}_{\perp}-\vec{\ell}_{\perp};\eta,\nu)G(\vec{\ell}_{\perp}-\vec{k}'_{\perp};\nu)^{(0)}\\
&=& \int d^{2}\ell_{\perp} {\cal Z}^{-1}(\vec{k}_{\perp}-\vec{\ell}_{\perp};\eta,\nu)\delta^{(2)}(\vec{\ell}_{\perp}-\vec{k}'_{\perp})
\nonumber \\
&=& \delta^{(2)}(\vec{k}_{\perp}-\vec{k}'_{\perp})+\textrm{counterterms}\,,\nonumber
\end{eqnarray}
where the superscript $(0)$ indicates the matrix element of the bare operator. Inverting the above equation leads to 
\begin{equation}
\label{ct}
\delta^{(2)}(\vec{k}_{\perp}-\vec{k}'_{\perp})=\int d^{2}\ell_{\perp} {\cal Z}(\vec{k}_{\perp}-\vec{\ell}_{\perp};\eta,\nu)G(\vec{\ell}_{\perp}-\vec{k}'_{\perp};\nu)
\,.
\end{equation}
The rapidity divergence term in Eq.~(\ref{rapdivterms}) is cancelled by setting 
\begin{eqnarray}
{\cal Z}(\vec{k}_{\perp}-\vec{\ell}_{\perp};\eta,\nu)&=&\delta^{(2)}(\vec{k}_{\perp}-\vec{\ell}_{\perp})-\bigg(\frac{\alpha_{s} N_{c}}{\pi^{2}}\bigg)\frac{w(\nu)^{2}}{\eta}\bigg[ \frac{1}{(\vec{k}_{\perp}-\vec{\ell}_{\perp})^{2}}\\
&&\qquad\qquad\qquad\qquad\qquad\qquad -\frac{1}{2}\delta^{(2)}(\vec{k}_{\perp}-\vec{\ell}_{\perp})\int \frac{d^{2}q_{\perp}}{(\vec{q}_{\perp}-\vec{k}_{\perp})^{2}}\frac{\vec{k}^{2}_{\perp}}{ \vec{q}^{\,\,2}_{\perp}}\bigg]\,.\nonumber
\end{eqnarray}
Inserting this expression into Eq.~(\ref{ct}) we find
\begin{eqnarray}
\delta^{(2)}(\vec{k}_{\perp}-\vec{k}'_{\perp})&=&G(\vec{k}_{\perp}-\vec{k}'_{\perp};\nu)-\bigg(\frac{\alpha_{s} N_{c}}{\pi^{2}}\bigg)\frac{w(\nu)^{2}}{\eta}\bigg[ \int d^{2}q_{\perp} \frac{G(\vec{q}_{\perp}-\vec{k}'_{\perp};\nu)}{(\vec{q}_{\perp}-\vec{k}_{\perp})^{2}}\\
&&\qquad\qquad\qquad\qquad\qquad\qquad -\frac{1}{2}G(\vec{k}_{\perp}-\vec{k}'_{\perp};\nu)\int \frac{d^{2}q_{\perp}}{(\vec{q}_{\perp}-\vec{k}_{\perp})^{2}}\frac{\vec{k}^{2}_{\perp}}{ \vec{q}^{\,\,2}_{\perp}}\bigg]\,,\nonumber
\end{eqnarray}
which when used in Eq.~(\ref{rapdivterms}) gives
\begin{equation}
\label{cs}
\sigma=\frac{2 \alpha^{2}_{s} C_{F}}{N_{c}}  \int  \frac{d^{2}\vec{k}_{\perp}}{\vec{k}_{\perp}^{2}} \int \frac{d^{2} \vec{k}^{'}_{\perp}}{\vec{k}_{\perp}^{'2}}G(\vec{k}_{\perp}-\vec{k}'_{\perp};\nu) +\dots
\end{equation}
where the singular terms in $\eta$ cancel and the dots indicate NLO terms that do not vanish in the $\eta \to 0$ limit. The dependence of $G(\vec{k}_{\perp}-\vec{k}'_{\perp};\nu)$ on $\nu$ is given by the rapidity RGE
\begin{equation}
\label{rrge}
\frac{d}{d \ln \nu}G(\vec{k}_{\perp}-\vec{k}'_{\perp};\nu)= \int d^{2}\ell_{\perp} \gamma_{\nu}(\vec{k}'_{\perp}-\vec{\ell}_{\perp})
G(\vec{\ell}_{\perp}-\vec{k}'_{\perp};\nu)\,,
\end{equation}
where
the rapidity anomalous dimension is determined from
\begin{equation}
\gamma_{\nu}(\vec{k}_{\perp}-\vec{k}'_{\perp}) = \int d^{2}\ell_{\perp} {\cal Z}(\vec{\ell}_{\perp},-\vec{k}'_{\perp};\eta,\nu)^{-1}\frac{d}{d \ln \nu}{\cal Z}(\vec{k}_{\perp}-\vec{\ell}_{\perp};\eta,\nu)\,.
\end{equation}
Using
\begin{equation}
\frac{d}{d \ln \nu} = \frac{\partial}{\partial \ln \nu}-w(\nu)^{2} \eta \frac{\partial}{\partial w^{2}}
\end{equation}
we find the leading-log (LL) rapidity anomalous dimension
\begin{equation}
\gamma_{\nu}(\vec{k}_{\perp}-\vec{k}'_{\perp})= \bigg(\frac{\alpha_{s} N_{c}}{\pi^{2}}\bigg)\bigg[ \frac{1}{(\vec{k}_{\perp}-\vec{k}'_{\perp})^{2}}
 -\frac{1}{2}\delta^{(2)}(\vec{k}_{\perp}-\vec{k}'_{\perp})\int \frac{d^{2}q_{\perp}}{(\vec{q}_{\perp}-\vec{k}_{\perp})^{2}}\frac{\vec{k}^{2}_{\perp}}{ \vec{q}^{\,\,2}_{\perp}}\bigg]\,,
\end{equation}
where we set $w(\nu) =1$. 
Using this LL expression in Eq.~(\ref{rrge}) gives
\begin{equation}
\frac{d}{d \ln \nu}G(\vec{k}_{\perp}-\vec{k}'_{\perp};\nu) = \bigg(\frac{\alpha_{s} N_{c}}{\pi^{2}}\bigg)\int \frac{d^{2}q_{\perp}}{(\vec{q}_{\perp}-\vec{k}_{\perp})^{2}}\bigg[G(\vec{q}_{\perp}-\vec{k}'_{\perp};\nu)-\frac{\vec{k}^{2}_{\perp}}{2\vec{q}^{\,\,2}_{\perp}}G(\vec{k}_{\perp}-\vec{k}'_{\perp};\nu)\bigg]\,.
\end{equation}
This is the BFKL equation [compare to Eq.~(3.58) in Ref.~\cite{Kovchegov:2012mbw}]. It can be solved by expanding $G(\vec{k}_{\perp}-\vec{k}'_{\perp};\nu)$ in eigenfunctions
\begin{equation}
G(\vec{k}_{\perp}-\vec{k}'_{\perp};\nu)= \sum^{\infty}_{n=-\infty}\int^{a+i\infty}_{a-i\infty}\frac{d\gamma}{2 \pi i} C_{n,\gamma}(\nu)
|\vec{k}_{\perp}|^{2(\gamma-1)}|\vec{k}'_{\perp}|^{2(\gamma^{*}-1)}e^{in(\phi-\phi')}\,,
\end{equation}
running in rapidity from $\ln \nu_{i} \sim 0$ to $\ln\nu_{f} \sim \ln s$, and then taking the inverse transform~\cite{Kovchegov:2012mbw}. The last step can only be done approximately. For large $\ln \nu_{f}$ one finds
\begin{eqnarray}
G(\vec{k}_{\perp}-\vec{k}'_{\perp};s)&=&\\
&& \hspace{-30pt}\frac{1}{2 \pi^{2}|\vec{k}_{\perp}||\vec{k}'_{\perp}|}\sqrt{\frac{\pi^{2}}{14 \zeta(3) \alpha_{s}(\mu) N_{c}s}}
\textrm{Exp}\bigg[\frac{4 \alpha_{s}(\mu) N_{c}}{\pi}\ln 2 \ln s-\frac{\pi \ln^{2}(|\vec{k}_{\perp}|/|\vec{k}'_{\perp}|}{14 \zeta(3)  \alpha_{s}(\mu) N_{c}s}\bigg]\,, \nonumber
\end{eqnarray}
with the leading piece being the first term in the exponent. Using this result in Eq.~(\ref{cs}) gives a cross section that has Regge behavior since it grows as a power of $s$
\begin{equation}
\sigma \sim s^{\alpha_{p}-1}\hspace{50pt} \alpha_{p}=1+\frac{4 \alpha_{s}(\mu) N_{c}}{\pi}\ln 2\,,
\end{equation}
where $\alpha_{p}$ is the pomeron intercept. 

Two important issues confront SCET: the role of Glauber interactions in factorization and the emergence of Regge behavior. Here, we have shown that at the perturbative level these two issues are actually the same. Glauber interactions between $n$ and $\bar{n}$ collinear quarks exhibit large rapidity logarithms at NLO in perturbation theory, and the rapidity RGE that resumes these large rapidity logarithms is the LL BFKL equation which gives rise to Regge behavior. While this is not the first attempt at incorporating Glauber interactions into SCET, it is the first time Glauber interactions in SCET have been connected to the emergence of Regge behavior in the theory. Clearly, the analysis presented here is based on a perturbative approach, and can only be considered a first (small) step towards a comprehensive treatment of Glauber interactions and hence Regge behavior in SCET. The ultimate goal is to include Glauber interactions to all orders so that conclusions about the role of Glauber interactions in factorization can be made to all orders, and that Regge behavior can be understood at the non-perturbative level. Both of these goals remain open questions in strong interactions, though a considerable effort has been devoted to them (especially the question of  non-perturbative Regge behavior~\cite{Balitsky:2001gj,Kovchegov:2012mbw}). The hope is that reformulating these questions in an effective field theory language will allow us to bring new tools to bear and arrive at a solution.

\acknowledgments
I would like to thank A. Stasto and T. Mehen for helpful discussions. This work was supported in part by the Director, Office of Science, Office of Nuclear Physics, of the U.S. Department of Energy under grant number DE-FG02-04ER41338. 


\bibliography{SCETBFKL}

\begin{thebibliography}{31}
\expandafter\ifx\csname natexlab\endcsname\relax\def\natexlab#1{#1}\fi
\expandafter\ifx\csname bibnamefont\endcsname\relax
  \def\bibnamefont#1{#1}\fi
\expandafter\ifx\csname bibfnamefont\endcsname\relax
  \def\bibfnamefont#1{#1}\fi
\expandafter\ifx\csname citenamefont\endcsname\relax
  \def\citenamefont#1{#1}\fi
\expandafter\ifx\csname url\endcsname\relax
  \def\url#1{\texttt{#1}}\fi
\expandafter\ifx\csname urlprefix\endcsname\relax\def\urlprefix{URL }\fi
\providecommand{\bibinfo}[2]{#2}
\providecommand{\eprint}[2][]{\url{#2}}

\bibitem[{\citenamefont{Collins and Sterman}(1981)}]{Collins:1981ta}
\bibinfo{author}{\bibfnamefont{J.~C.} \bibnamefont{Collins}} \bibnamefont{and}
  \bibinfo{author}{\bibfnamefont{G.~F.} \bibnamefont{Sterman}},
  \bibinfo{journal}{Nucl. Phys. B} \textbf{\bibinfo{volume}{185}},
  \bibinfo{pages}{172} (\bibinfo{year}{1981}).

\bibitem[{\citenamefont{Collins and Soper}(1981)}]{Collins:1981uk}
\bibinfo{author}{\bibfnamefont{J.~C.} \bibnamefont{Collins}} \bibnamefont{and}
  \bibinfo{author}{\bibfnamefont{D.~E.} \bibnamefont{Soper}},
  \bibinfo{journal}{Nucl. Phys. B} \textbf{\bibinfo{volume}{193}},
  \bibinfo{pages}{381} (\bibinfo{year}{1981}).

\bibitem[{\citenamefont{Collins and Soper}(1987)}]{Collins:1987pm}
\bibinfo{author}{\bibfnamefont{J.~C.} \bibnamefont{Collins}} \bibnamefont{and}
  \bibinfo{author}{\bibfnamefont{D.~E.} \bibnamefont{Soper}},
  \bibinfo{journal}{Ann. Rev. Nucl. Part. Sci.} \textbf{\bibinfo{volume}{37}},
  \bibinfo{pages}{383} (\bibinfo{year}{1987}).

\bibitem[{\citenamefont{Collins et~al.}(1988)\citenamefont{Collins, Soper, and
  Sterman}}]{Collins:1989gx}
\bibinfo{author}{\bibfnamefont{J.~C.} \bibnamefont{Collins}},
  \bibinfo{author}{\bibfnamefont{D.~E.} \bibnamefont{Soper}}, \bibnamefont{and}
  \bibinfo{author}{\bibfnamefont{G.~F.} \bibnamefont{Sterman}},
  \emph{\bibinfo{title}{Perturbative Quantum Chromodynamics}}
  (\bibinfo{publisher}{World Scientific Pub}, \bibinfo{year}{1988}),
  vol.~\bibinfo{volume}{5}, chap.~\bibinfo{chapter}{1}, pp.
  \bibinfo{pages}{1--91}, \eprint{arXiv: hep-ph/0409313}.

\bibitem[{\citenamefont{Landau}(1959)}]{Landau:1959fi}
\bibinfo{author}{\bibfnamefont{L.}~\bibnamefont{Landau}},
  \bibinfo{journal}{Nucl. Phys.} \textbf{\bibinfo{volume}{13}},
  \bibinfo{pages}{181} (\bibinfo{year}{1959}).

\bibitem[{\citenamefont{Coleman and Norton}(1965)}]{Coleman:1965xm}
\bibinfo{author}{\bibfnamefont{S.}~\bibnamefont{Coleman}} \bibnamefont{and}
  \bibinfo{author}{\bibfnamefont{R.}~\bibnamefont{Norton}},
  \bibinfo{journal}{Nuovo Cim.} \textbf{\bibinfo{volume}{38}},
  \bibinfo{pages}{438} (\bibinfo{year}{1965}).

\bibitem[{\citenamefont{Bodwin et~al.}(1981)\citenamefont{Bodwin, Brodsky, and
  Lepage}}]{Bodwin:1981fv}
\bibinfo{author}{\bibfnamefont{G.~T.} \bibnamefont{Bodwin}},
  \bibinfo{author}{\bibfnamefont{S.~J.} \bibnamefont{Brodsky}},
  \bibnamefont{and} \bibinfo{author}{\bibfnamefont{G.~P.}
  \bibnamefont{Lepage}}, \bibinfo{journal}{Phys. Rev. Lett.}
  \textbf{\bibinfo{volume}{47}}, \bibinfo{pages}{1799} (\bibinfo{year}{1981}).

\bibitem[{\citenamefont{Collins et~al.}(1982)\citenamefont{Collins, Soper, and
  Sterman}}]{Collins:1981tt}
\bibinfo{author}{\bibfnamefont{J.~C.} \bibnamefont{Collins}},
  \bibinfo{author}{\bibfnamefont{D.~E.} \bibnamefont{Soper}}, \bibnamefont{and}
  \bibinfo{author}{\bibfnamefont{G.~F.} \bibnamefont{Sterman}},
  \bibinfo{journal}{Phys. Lett. B} \textbf{\bibinfo{volume}{109}},
  \bibinfo{pages}{388} (\bibinfo{year}{1982}).

\bibitem[{\citenamefont{Bodwin}(1985)}]{Bodwin:1984hc}
\bibinfo{author}{\bibfnamefont{G.~T.} \bibnamefont{Bodwin}},
  \bibinfo{journal}{Phys. Rev. D} \textbf{\bibinfo{volume}{31}},
  \bibinfo{pages}{2616} (\bibinfo{year}{1985}), \bibinfo{note}{[Erratum: Phys.
  Rev. D {\bf 34} (Dec. 1986), 3932]}.

\bibitem[{\citenamefont{Collins et~al.}(1985)\citenamefont{Collins, Soper, and
  Sterman}}]{Collins:1985ue}
\bibinfo{author}{\bibfnamefont{J.~C.} \bibnamefont{Collins}},
  \bibinfo{author}{\bibfnamefont{D.~E.} \bibnamefont{Soper}}, \bibnamefont{and}
  \bibinfo{author}{\bibfnamefont{G.~F.} \bibnamefont{Sterman}},
  \bibinfo{journal}{Nucl. Phys. B} \textbf{\bibinfo{volume}{261}},
  \bibinfo{pages}{104} (\bibinfo{year}{1985}).

\bibitem[{\citenamefont{Aybat and Sterman}(2009)}]{Aybat:2008ct}
\bibinfo{author}{\bibfnamefont{S.~M.} \bibnamefont{Aybat}} \bibnamefont{and}
  \bibinfo{author}{\bibfnamefont{G.~F.} \bibnamefont{Sterman}},
  \bibinfo{journal}{Phys. Lett. B} \textbf{\bibinfo{volume}{671}},
  \bibinfo{pages}{46} (\bibinfo{year}{2009}), \eprint{arXiv:0811.0246}.

\bibitem[{\citenamefont{Bauer et~al.}(2000)\citenamefont{Bauer, Fleming, and
  Luke}}]{Bauer:2000ew}
\bibinfo{author}{\bibfnamefont{C.~W.} \bibnamefont{Bauer}},
  \bibinfo{author}{\bibfnamefont{S.}~\bibnamefont{Fleming}}, \bibnamefont{and}
  \bibinfo{author}{\bibfnamefont{M.~E.} \bibnamefont{Luke}},
  \bibinfo{journal}{Phys. Rev. D} \textbf{\bibinfo{volume}{63}},
  \bibinfo{pages}{014006} (\bibinfo{year}{2000}), \eprint{hep-ph/0005275}.

\bibitem[{\citenamefont{Bauer et~al.}(2001)\citenamefont{Bauer, Fleming,
  Pirjol, and Stewart}}]{Bauer:2000yr}
\bibinfo{author}{\bibfnamefont{C.~W.} \bibnamefont{Bauer}},
  \bibinfo{author}{\bibfnamefont{S.}~\bibnamefont{Fleming}},
  \bibinfo{author}{\bibfnamefont{D.}~\bibnamefont{Pirjol}}, \bibnamefont{and}
  \bibinfo{author}{\bibfnamefont{I.~W.} \bibnamefont{Stewart}},
  \bibinfo{journal}{Phys. Rev. D} \textbf{\bibinfo{volume}{63}},
  \bibinfo{pages}{114020} (\bibinfo{year}{2001}), \eprint{hep-ph/0011336}.

\bibitem[{\citenamefont{Bauer and Stewart}(2001)}]{Bauer:2001ct}
\bibinfo{author}{\bibfnamefont{C.~W.} \bibnamefont{Bauer}} \bibnamefont{and}
  \bibinfo{author}{\bibfnamefont{I.~W.} \bibnamefont{Stewart}},
  \bibinfo{journal}{Phys. Lett. B} \textbf{\bibinfo{volume}{516}},
  \bibinfo{pages}{134} (\bibinfo{year}{2001}), \eprint{hep-ph/0107001}.

\bibitem[{\citenamefont{Bauer et~al.}(2002{\natexlab{a}})\citenamefont{Bauer,
  Pirjol, and Stewart}}]{Bauer:2001yt}
\bibinfo{author}{\bibfnamefont{C.~W.} \bibnamefont{Bauer}},
  \bibinfo{author}{\bibfnamefont{D.}~\bibnamefont{Pirjol}}, \bibnamefont{and}
  \bibinfo{author}{\bibfnamefont{I.~W.} \bibnamefont{Stewart}},
  \bibinfo{journal}{Phys. Rev. D} \textbf{\bibinfo{volume}{65}},
  \bibinfo{pages}{054022} (\bibinfo{year}{2002}{\natexlab{a}}),
  \eprint{hep-ph/0109045}.

\bibitem[{\citenamefont{Bauer et~al.}(2002{\natexlab{b}})\citenamefont{Bauer,
  Fleming, Pirjol, Rothstein, and Stewart}}]{Bauer:2002nz}
\bibinfo{author}{\bibfnamefont{C.~W.} \bibnamefont{Bauer}},
  \bibinfo{author}{\bibfnamefont{S.}~\bibnamefont{Fleming}},
  \bibinfo{author}{\bibfnamefont{D.}~\bibnamefont{Pirjol}},
  \bibinfo{author}{\bibfnamefont{I.~Z.} \bibnamefont{Rothstein}},
  \bibnamefont{and} \bibinfo{author}{\bibfnamefont{I.~W.}
  \bibnamefont{Stewart}}, \bibinfo{journal}{Phys. Rev. D}
  \textbf{\bibinfo{volume}{66}}, \bibinfo{pages}{014017}
  (\bibinfo{year}{2002}{\natexlab{b}}), \eprint{hep-ph/0202088}.

\bibitem[{\citenamefont{Liu and Ma}(2008)}]{Liu:2008cc}
\bibinfo{author}{\bibfnamefont{F.}~\bibnamefont{Liu}} \bibnamefont{and}
  \bibinfo{author}{\bibfnamefont{J.}~\bibnamefont{Ma}} (\bibinfo{year}{2008}),
  \eprint{arXiv:0802.2973}.

\bibitem[{\citenamefont{Bauer et~al.}(2011)\citenamefont{Bauer, Lange, and
  Ovanesyan}}]{Bauer:2010cc}
\bibinfo{author}{\bibfnamefont{C.~W.} \bibnamefont{Bauer}},
  \bibinfo{author}{\bibfnamefont{B.~O.} \bibnamefont{Lange}}, \bibnamefont{and}
  \bibinfo{author}{\bibfnamefont{G.}~\bibnamefont{Ovanesyan}},
  \bibinfo{journal}{JHEP} \textbf{\bibinfo{volume}{1107}}, \bibinfo{pages}{077}
  (\bibinfo{year}{2011}), \eprint{arXiv:1010.1027}.

\bibitem[{\citenamefont{Idilbi and Majumder}(2009)}]{Idilbi:2008vm}
\bibinfo{author}{\bibfnamefont{A.}~\bibnamefont{Idilbi}} \bibnamefont{and}
  \bibinfo{author}{\bibfnamefont{A.}~\bibnamefont{Majumder}},
  \bibinfo{journal}{Phys. Rev. D} \textbf{\bibinfo{volume}{80}},
  \bibinfo{pages}{054022} (\bibinfo{year}{2009}), \eprint{arXiv:0808.1087}.

\bibitem[{\citenamefont{D'Eramo et~al.}(2011)\citenamefont{D'Eramo, Liu, and
  Rajagopal}}]{DEramo:2010ak}
\bibinfo{author}{\bibfnamefont{F.}~\bibnamefont{D'Eramo}},
  \bibinfo{author}{\bibfnamefont{H.}~\bibnamefont{Liu}}, \bibnamefont{and}
  \bibinfo{author}{\bibfnamefont{K.}~\bibnamefont{Rajagopal}},
  \bibinfo{journal}{Phys. Rev. D} \textbf{\bibinfo{volume}{84}},
  \bibinfo{pages}{065015} (\bibinfo{year}{2011}), \eprint{arXiv:1006.1367}.

\bibitem[{\citenamefont{Benzke et~al.}(2013)\citenamefont{Benzke, Brambilla,
  Escobedo, and Vairo}}]{Benzke:2012sz}
\bibinfo{author}{\bibfnamefont{M.}~\bibnamefont{Benzke}},
  \bibinfo{author}{\bibfnamefont{N.}~\bibnamefont{Brambilla}},
  \bibinfo{author}{\bibfnamefont{M.~A.} \bibnamefont{Escobedo}},
  \bibnamefont{and} \bibinfo{author}{\bibfnamefont{A.}~\bibnamefont{Vairo}},
  \bibinfo{journal}{JHEP} \textbf{\bibinfo{volume}{1302}}, \bibinfo{pages}{129}
  (\bibinfo{year}{2013}), \eprint{arXiv:1208.4253}.

\bibitem[{\citenamefont{Stewart and Rothstein}({\natexlab{a}})}]{iainSCET10}
\bibinfo{author}{\bibfnamefont{I.~W.} \bibnamefont{Stewart}} \bibnamefont{and}
  \bibinfo{author}{\bibfnamefont{I.~Z.} \bibnamefont{Rothstein}},
  \bibinfo{note}{{(work in progress). Talk ``Glauber Gluons in SCET'' presented
  by I. W. Stewart at SCET2010 in Ringberg, Germany}},
  \urlprefix\url{https://indico.mpp.mpg.de/conferenceDisplay.py?confId=632}.

\bibitem[{\citenamefont{Stewart and Rothstein}({\natexlab{b}})}]{rothsteinKITP}
\bibinfo{author}{\bibfnamefont{I.~W.} \bibnamefont{Stewart}} \bibnamefont{and}
  \bibinfo{author}{\bibfnamefont{I.~Z.} \bibnamefont{Rothstein}},
  \bibinfo{note}{{(work in progress). Talk ``Progress in EFT'' presented by I.
  Z. Rothstein at Quantum Fields Beyond Perturbation Theory KITP, Santa
  Barbara}}, \urlprefix\url{http://online.kitp.ucsb.edu/online/qft-c14/}.

\bibitem[{\citenamefont{Donoghue and Wyler}(2010)}]{Donoghue:2009cq}
\bibinfo{author}{\bibfnamefont{J.~F.} \bibnamefont{Donoghue}} \bibnamefont{and}
  \bibinfo{author}{\bibfnamefont{D.}~\bibnamefont{Wyler}},
  \bibinfo{journal}{Phys. Rev. D} \textbf{\bibinfo{volume}{81}},
  \bibinfo{pages}{114023} (\bibinfo{year}{2010}), \eprint{arXiv:0908.4559}.

\bibitem[{\citenamefont{Donoghue}(2009)}]{Donoghue:2009mn}
\bibinfo{author}{\bibfnamefont{J.~F.} \bibnamefont{Donoghue}},
  \bibinfo{journal}{PoS} \textbf{\bibinfo{volume}{EFT09}}, \bibinfo{pages}{001}
  (\bibinfo{year}{2009}), \eprint{arXiv:0909.0021}.

\bibitem[{\citenamefont{Fadin et~al.}(1975)\citenamefont{Fadin, Kuraev, and
  Lipatov}}]{Fadin:1975cb}
\bibinfo{author}{\bibfnamefont{V.~S.} \bibnamefont{Fadin}},
  \bibinfo{author}{\bibfnamefont{E.}~\bibnamefont{Kuraev}}, \bibnamefont{and}
  \bibinfo{author}{\bibfnamefont{L.}~\bibnamefont{Lipatov}},
  \bibinfo{journal}{Phys. Lett. B} \textbf{\bibinfo{volume}{60}},
  \bibinfo{pages}{50} (\bibinfo{year}{1975}).

\bibitem[{\citenamefont{Balitsky and Lipatov}(1978)}]{Balitsky:1978ic}
\bibinfo{author}{\bibfnamefont{I.}~\bibnamefont{Balitsky}} \bibnamefont{and}
  \bibinfo{author}{\bibfnamefont{L.}~\bibnamefont{Lipatov}},
  \bibinfo{journal}{Sov. J. Nucl. Phys.} \textbf{\bibinfo{volume}{28}},
  \bibinfo{pages}{822} (\bibinfo{year}{1978}).

\bibitem[{\citenamefont{Kovchegov and Levin}(2012)}]{Kovchegov:2012mbw}
\bibinfo{author}{\bibfnamefont{Y.~V.} \bibnamefont{Kovchegov}}
  \bibnamefont{and} \bibinfo{author}{\bibfnamefont{E.}~\bibnamefont{Levin}},
  \emph{\bibinfo{title}{{Quantum chromodynamics at high energy}}}
  (\bibinfo{publisher}{Cambridge University Press}, \bibinfo{address}{The
  Edinburgh Building, Cambridge CB2 8RU, UK}, \bibinfo{year}{2012}),
  \bibinfo{edition}{1st} ed.

\bibitem[{\citenamefont{Chiu et~al.}(2012{\natexlab{a}})\citenamefont{Chiu,
  Jain, Neill, and Rothstein}}]{Chiu:2011qc}
\bibinfo{author}{\bibfnamefont{J.-y.} \bibnamefont{Chiu}},
  \bibinfo{author}{\bibfnamefont{A.}~\bibnamefont{Jain}},
  \bibinfo{author}{\bibfnamefont{D.}~\bibnamefont{Neill}}, \bibnamefont{and}
  \bibinfo{author}{\bibfnamefont{I.~Z.} \bibnamefont{Rothstein}},
  \bibinfo{journal}{Phys. Rev. Lett.} \textbf{\bibinfo{volume}{108}},
  \bibinfo{pages}{151601} (\bibinfo{year}{2012}{\natexlab{a}}),
  \eprint{arXiv:1104.0881}.

\bibitem[{\citenamefont{Chiu et~al.}(2012{\natexlab{b}})\citenamefont{Chiu,
  Jain, Neill, and Rothstein}}]{Chiu:2012ir}
\bibinfo{author}{\bibfnamefont{J.-Y.} \bibnamefont{Chiu}},
  \bibinfo{author}{\bibfnamefont{A.}~\bibnamefont{Jain}},
  \bibinfo{author}{\bibfnamefont{D.}~\bibnamefont{Neill}}, \bibnamefont{and}
  \bibinfo{author}{\bibfnamefont{I.~Z.} \bibnamefont{Rothstein}},
  \bibinfo{journal}{JHEP} \textbf{\bibinfo{volume}{1205}}, \bibinfo{pages}{084}
  (\bibinfo{year}{2012}{\natexlab{b}}), \eprint{arXiv:1202.0814}.

\bibitem[{\citenamefont{Balitsky}(2001)}]{Balitsky:2001gj}
\bibinfo{author}{\bibfnamefont{I.}~\bibnamefont{Balitsky}}
  (\bibinfo{year}{2001}), \eprint{hep-ph/0101042}.

\end{thebibliography}

\end{document}